# Modulation of the occurrence of heat waves over the Euro-Mediterranean region by the intensity of the Atlantic Multidecadal Variability

Saïd Qasmi[1,2], Emilia Sanchez-Gomez[2], Yohan Ruprich-Robert[3], Julien Boé[2] and Christophe Cassou[2]

[1] *CNRM, Université de Toulouse, Météo-France, CNRS, Toulouse, France*
[2] *CECI, Université de Toulouse, CNRS, Cerfacs, Toulouse, France*
[3] *Barcelona Supercomputing Center, Barcelona, Spain*

*Corresponding author address:* Saïd Qasmi, Centre National de Recherches Météorologiques, 42 Avenue Gaspard Coriolis, 31057 Toulouse, France.
*E-mail: qasmi@cerfacs.fr*

# ABSTRACT

The influence of the Atlantic Multidecadal Variability (AMV) and its amplitude on the Euro-Mediterranean summer climate is studied in two climate models, namely CNRM-CM5 and EC-Earth3P. Large ensembles of idealized experiments have been conducted in which North Atlantic sea surface temperatures (SSTs) are relaxed towards different amplitudes of the observed AMV anomalies. In agreement with observations, during a positive phase of the AMV both models simulate an increase (decrease) of temperature and a decrease (increase) of precipitation over the Mediterranean basin (northern half of Europe). An increase in the number of heat wave (HW) days is also found over the continental Mediterranean region. In terms of atmospheric circulation, an anomalous ridge (trough) over the Mediterranean basin (northern Europe) induces a thermodynamical response explaining the increase in the occurrence of HW. The anticyclonic anomaly over the Mediterranean basin is associated with drier soils and a reduction of cloud cover, which concomitantly induce a decrease (increase) of the latent (sensible) heat flux and an increase of the downward radiative fluxes over lands. The source mechanisms leading to the mid-troposphere anomalies over North Atlantic and its adjacent continents are discussed. It is found that both the tropical and extra-tropical parts of the AMV modulate the atmospheric circulation over the Euro-Atlantic region. The number of HW days increases linearly with the amplitude of the AMV. However, the strength of this relationship differs between the models, and depends on their intrinsic biases, raising questions regarding the robustness of the mechanisms of teleconnections associated with the AMV.

# 1. Introduction

Heat waves over Europe are associated with major damages in many societal areas through increased mortality (Robine et al. 2008; Guo et al. 2017), decreased crop production (Ciais et al. 2005; Loboda et al. 2017), increased droughts (Zampieri et al. 2009). Anticipating such extreme weather events, especially their probability of occurrence has the potential to limit their harmful impacts.

Heat waves (hereafter HW) are primarily driven by internal atmospheric variability (Dole et al. 2011), making therefore challenging their predictability on seasonal timescales and even more on decadal timescales (Hanlon et al. 2013). Yet, HW properties can be modulated by boundary conditions of the climate system, like anthropogenic forcings (Hansen et al. 2012), but also by several of its components, *e.g.* the surface via soil moisture (Alexander 2011), the ocean through sea surface temperature (SST) (Alexander et al. 2009, Ruprich-Robert et al. 2018). At decadal timescales, oceanic modes of variability such as the Atlantic Multidecadal Variability (AMV) (Knight et al. 2005) seem to modulate the HW variations over the adjacent continents, *e.g.* the United Kingdom (UK), where several station-based HW measurements show a decadal variability during the 20th century (Sanderson et al. 2017).

At decadal timescales, prediction skill of state of the art numerical forecasts mainly comes from anthropogenic external forcing (van Oldenborgh et al. 2012). Additional skill comes from the initialization to observations of the different components of the simulated climate system, which aims to phase numerical decadal predictions variability with the observed one. In particular, decadal predictions are skillful in predicting the North Atlantic and the AMV several years in advance thanks to the initialization process (Robson et al. 2012, Yeager et al. 2018, García-Serrano et al. 2012). Yet, mean climate predictability over the adjacent continents (beyond ~1 yr) is rapidly lost (Doblas-Reyes et al. 2013). Similar results are also found for initialized predictions of climate extremes (Khodayar et al. 2015, Seager and Ting 2017, Liu et al. 2019).

Such a loss of predictability seems somehow paradoxical given the apparent links that have been documented between the AMV and summer climate over Europe both in models and observations. Sutton and Hodson (2005) show that during a positive phase of the AMV, warmer conditions are obtained over central Europe, particularly over the Mediterranean basin. Concomitantly, a decrease in precipitation is obtained over this region while an increase is observed over the northern half of Europe (Sutton and Dong 2012). Mariotti and Dell'Aquila (2011) also found that about 30% of summer temperature anomalies over the Mediterranean basin are explained by the AMV.

Current climate models show uneven results in the simulation of the teleconnection between the AMV and summer temperature over Europe, both in historical and pre-industrial control CMIP5 simulations (Qasmi et al. 2017). A large intermodel spread exists in the level of teleconnectivity, partly because of numerous uncertainties in the intrinsic properties of the AMV such as its amplitude, frequency, and in the coupling mechanisms between the ocean and the atmosphere. A key challenge for the climate research community is to overcome these deficiencies, by understanding how the ocean decadal variability actually influences climate over lands and how this information can be used for valuable climate predictions.

Several mechanisms have been proposed to explain the relationship between the AMV and summer climate variability over Europe. Negative sea level pressure (SLP) anomalies are observed over the British Isles during a positive phase of the AMV (hereafter AMV+) (Sutton and Hodson 2005; Sutton and Dong 2012; Ting et al. 2014), associated with positive precipitation anomalies and the absence of a significant temperature signal over Northern Europe. These negative SLP anomalies have been related to the southern lobe of the summer North Atlantic Oscillation (NAO) in its negative phase (hereafter NAO-). The latter seems to be more excited during the AMV+ phase (Mariotti and Dell'Aquila 2011, Linderholm and Folland 2017), with an influence extending to the Mediterranean, as suggested by Mariotti and Dell'Aquila (2011) who found that 30% of the decadal variance of Mediterranean summer precipitations is explained by the NAO. O'Reilly et al. (2017) confirm this result by separating the dynamic and thermodynamic influences of the AMV, mentioning that temperature anomalies over Mediterranean are partly explained by the large-scale atmospheric circulation in the Euro-Atlantic region. Using dedicated experiments, in which North Atlantic SSTs are prescribed, Sutton and Hodson (2007) show that the SLP response to an AMV forcing is consistent with the summer NAO- phase. This is confirmed by Ruprich-Robert et al. (2017) who also find a similar response in SLP and significant precipitation and temperature anomalies over Europe using idealized AMV coupled simulations.

Impacts of the AMV over the Euro-Mediterranean region have been documented in terms of mean climate, both in observations and models, but this is not the case for extreme events. Yet, the Mediterranean basin is considered as a climate change hotspot (Giorgi 2006) for which an assessment of the risks related to climate change is important since HW are expected to be more frequent in the next decades (Meehl and Tebaldi 2004; Diffenbaugh et al. 2007).

In this context, the Decadal Climate Prediction Project (DCPP) endorsed by CMIP6 (Eyring et al. 2016), and the European H2020 PRIMAVERA project (PRocess-based climate sIMulation: AdVances in high-resolution modeling and European climate Risk Assessment, https://www.primavera-h2020.eu/) aim to improve the understanding of the processes linking the AMV and climate decadal variability. For this purpose, a coordinated experimental protocol using partial coupling experiments has been proposed. In these simulations, North Atlantic SSTs are restored towards anomalies representative of the observed AMV while the rest of the system evolves freely (Boer et al. 2016, DCPP-Component C).

This paper aims to assess the influence of the AMV and also its amplitude on the European climate, particularly on HW, by using DCPP-compliant experiments from two climate models. The experimental protocol is presented in section 2. The mean temperature and HW responses to the AMV forcing, the mechanisms of teleconnection between the AMV and European HW, as well as their sensitivity to the amplitude of the AMV are detailed in section 3. We discuss and conclude our results in section 4.

## 2. Methods and data

*a. Model sensitivity experiments*

In this study, two climate models are used: EC-Earth3P (Haarsma et al. 2019) and CNRM-CM5 (Voldoire et al. 2013) from the CMIP6 and CMIP5 ensemble, respectively. The DCPP framework and the experimental protocol of the idealized AMV simulations are detailed by Boer et al. (2016) (Table C1, Component C1.2 and C1.3, see also Technical Note 1: https://www.wcrp-climate.org/wgsip/documents/Tech-Note-1.pdf). Fig. 1a shows the pattern of SST anomalies used for the SST restoring for an AMV+ phase. This is done by adding a feedback term to the non-solar total heat flux in the surface temperature equation (Haney 1971). The restoring coefficient is fixed at -40 W.m-2.K-1, which is equivalent to a two-month damping for a 50 m-deep mixed layer. In order to prevent a potential oceanic circulation drift introduced by the imposed SST anomalies, a restoring of sea surface salinity (SSS) through a freshwater flux correction is also applied in EC-Earth3P to conserve the North Atlantic surface density to a neutral state (see DCPP-C Technical Note 2: https://www.wcrp-climate.org/wgsip/documents/Tech-Note-2.pdf). For each model, an AMV+ and an AMV- ensemble of 25 members of 10 years are computed. The two ensembles differ only in the sign of the targeted SST anomalies. For an AMV+ phase, SST anomalies from Fig. 1a are superimposed on the model's own climatology to obtain the SST pattern towards which modeled SSTs are restored. The initial conditions for CNRM-CM5 (EC-Earth3P) for all the components are arbitrarily selected every 5 (6) years within a 125 (150) yr-long control simulation conducted with CNRM-CM5 (EC-Earth3P), in which radiative forcings are constant during integration, and are fixed to the 1985 (1950) estimated values. The control simulations are detailed by Oudar et al. (2017) and Haarsma et al. (2019) for CNRM-CM5 and EC-Earth3P, respectively. In order to assess the sensitivity of the atmospheric response to the amplitude of the AMV, two additional sets of ensembles are computed for each model by multiplying the amplitude of the targeted SST anomalies by 2 and 3. The two additional ensembles are termed 2xAMV and 3xAMV respectively, and the reference ensemble is referred to as 1xAMV.

Since SSTs are not imposed but relaxed in our experiments, the amplitude of the ensemble mean anomalies of the resulting SSTs after the restoring in CNRM-CM5 are always lower than the targeted SST for all amplitudes and both phases of the AMV, with an ensemble spread of about 0.4°C (Fig. 1b). Compared to CNRM-CM5, EC-Earth3P simulates a warmer North Atlantic SST mean state of ~0.8°C, explaining the warmer SST target for both AMV+ and AMV-. In addition, all SST ensemble means from EC-Earth3P are shifted in a way that for the AMV- phase, the SSTs are colder than the targeted SST; this shift is more pronounced for the AMV+ phase for which the SSTs are much colder than the targeted SSTs. This systematic shift is due to a technical feature related to the restoring protocol: a temporal linear interpolation of the targeted SST fields from a monthly timestep to the model's timestep results in an underestimation of the amplitude of the seasonal cycle in the model. This is apparent in EC-Earth3P whereas CNRM-CM5 seems to be less sensitive to this interpolation. Despite these discrepancies, the ensemble mean differences between AMV+ and AMV- are similar between the two models for each amplitude of AMV, indicating that the imposed anomalies are the same for both models.

In order to assess the respective contributions of the tropical and extra-tropical parts of the the AMV, additional twin ensemble experiments (also proposed in DCPP-C) where the AMV full pattern is split into tropical and extra-tropical anomalies have been conducted with CNRM-CM5 for the 1xAMV case. These experiments are respectively named 1xAMV$_T$ and 1xAMV$_E$.

The AMV-forced anomalies (also named response to the AMV) are defined as the ensemble mean differences between the AMV+ and AMV- phases. For readability reasons, only the responses for 2xAMV experiments are shown in the main manuscript, while the responses for 1xAMV and 3xAMV are shown in the Supplementary Material.

*b. Definition of heat waves and number of heat wave days*

The definition of a HW following Ruprich-Robert et al. (2018) is adopted in this paper. For a given amplitude of the AMV, and for each member of the AMV+ and AMV- ensemble, a HW is defined as a group of days that satisfy three criteria: (i) Tx must exceed T90 for at least three consecutive days, (ii) Tx averaged over the entire event must exceed T90, and iii) Tx for each day of the event must exceed the T75, where Tx is the daily maximum 2-m air temperature, and T90 (T75) corresponds to the 90th (70th) percentile of the Tx distribution built from the all the members of the AMV+ and AMV- experiments during the June-July-August (JJA) period. The number of HW days corresponds to the number of days during summer that meet the HW criteria.

## 3. Results

*a. Impact of the AMV on the European summer climate*

Consistent with previous studies based on models (Ruprich-Robert et al. 2017) and observations (O'Reilly et al. 2017), both models simulate in 2xAMV a near surface warming over the Mediterranean basin and northern Scandinavia in JJA, with a 2-meter temperature (T2m) response of ~0.5°C (Fig. 2ab). This warming extends to the British Isles, Northern France and Central Europe in EC-Earth3P. Conversely, no signal is detectable over these regions in CNRM-CM5; a cooling is even obtained over the Baltic region in the latter. This difference between the two models can be explained by discrepancies between the AMV-forced responses through the atmospheric circulation and/or in the thermodynamically-based processes (detailed in section 3).

A dipole of precipitation anomalies is obtained over Europe in both models, with negative (positive) anomalies South (North) of 45°N, with significant values over Scandinavia and the North Sea coasts, except for EC-Earth3P, in which, following the T2m response, drier conditions are also found over western Europe. For both precipitation and T2m, continent-scale anomalies in 3xAMV have the same spatial structure compared to 2xAMV, but with a greater amplitude (Fig. 3a-d), while they are reduced in the 1xAMV experiments, in which the signals are hardly significant (Fig. 4a-d).

In consistence with previous studies (Sutton and Hodson 2007, Sutton and Dong 2012), an anomalous trough is obtained over North Sea and Scandinavia in both models in 2xAMV (black contours in Fig. 2ef), coherent with the increase of precipitation over these regions. Concomitantly, an anomalous ridge at 500 hPa above negative SLP anomalies is also found over the Mediterranean region. This ridge-trough meridional dipole contributes to explain the precipitation and T2m responses on each side of 45°N over Europe. Noticeably, SLP and geopotential at 500 hPa (Z500) anomalies both are stronger and cover a larger area in CNRM-CM5 than in EC-Earth3P, with a minimum over the west of the UK, explaining the above-mentioned differences between the two models in terms of precipitation and temperature over western Europe. The dipolar circulation anomalies are also found in

3xAMV but with a stronger amplitude (Fig. 3ef), whereas it is less robust in 1xAMV, in which no significant anomalies of SLP or Z500 are detected (Fig. 4ef).

An increase in the number of HW days is obtained in both models over the Mediterranean basin in 2xAMV (Fig. 2gh). However, the location of the maximum anomalies differs between the two models: Anatolia, the Levant and Maghreb for EC-Earth3P, while Greece, Italy and Turkey are more impacted in CNRM-CM5. For both models, the number of HW days per summer over these regions is increased by ~20% in average (up to 50% over the eastern Mediterranean) relative to the climatological number of HW days in 2xAMV (~7 per summer, black contours in Fig. 2gh). Coherently with the mean climate responses in T2m and precipitations, positive HW anomalies are also found in 1xAMV (3xAMV), with an increase of ~10% (~30%).

In the following section, we investigate the mechanisms that may explain the climate responses characterized in this section. We will focus both on processes leading to local thermodynamic changes and those causing the atmospheric circulation changes.

*b. Influence of the AMV-forced SST anomalies on thermodynamical processes*

The Z500 anticyclonic anomalies located over the Mediterranean region are associated with a significant decrease of total cloud cover in EC-Earth3P in 2xAMV (Fig. 5a). This decrease is less pronounced in CNRM-CM5 (Fig. 5b), and restricted to the eastern part of the Mediterranean basin (a slight increase is even obtained over the western Maghreb, Iberian Peninsula, and northern Europe). These anomalies induce changes in the surface radiation budget: positive anomalies in the sum of the downward longwave (LW) and shortwave (SW) radiation are found over regions associated with a decrease of total cloud cover (Fig. 5cd), which is spatially consistent with the T2m response (Fig. 2ab) and the increase of the number of HW (Fig. 2gh). Note that in CNRM-CM5, positive downward LW+SW radiation anomalies are also obtained over northwestern African coasts and the Iberian Peninsula although cloud cover increases over these regions, suggesting the impact of other processes on the positive downward radiation. The lower troposphere warming and moistening also impact downward LW radiation, and partly explain the positive LW+SW anomalies over the Mediterranean basin. Unfortunately, LW and SW radiation fields in clear-sky conditions have not been saved to precisely assess the radiative contribution of cloud cover anomalies.

In both models, a negative latent heat flux response, *i.e.* a decrease in evapotranspiration, is obtained over the eastern Mediterranean coasts (Fig. 5ef), where the increase of the number of HW days is the largest (Fig. 2gh). As no particular drying is obtained over these regions in the previous spring and winter (not shown), this decrease of evapotranspiration is associated with the concomitant negative precipitation response (Fig. 2cd) therefore causing a surface heating and enhancing the probability of HW occurrence. Positive anomalies of sensible heat flux (SH) are obtained over the Mediterranean coasts, particularly over the eastern side of the basin, because of warmer soils than the lower atmosphere, which are directly impacted by the increase in downward SW radiation and by the drier soil conditions (Fig. 5gh).

In addition to the direct effect of the negative cloud cover anomalies on the radiative fluxes, other processes may contribute to the T2m and HW responses, especially over the western Mediterranean and the Levant, where a response of turbulent heat fluxes is not detectable. In

CNRM-CM5, it is found that the northwestern African coasts, the Iberian Peninsula and the Levant are under the influence of the advection by the climatological wind of positive temperatures anomalies from North Atlantic and the Mediterranean sea, where SST anomalies are positive during an AMV+ phase, contributing to the increase obtained in the downward LW fluxes (not shown).

Coherently with the T2m and HW responses, similar responses in the radiative fluxes are found in the 3xAMV and 1xAMV experiments (Fig. 6-7), in which the amplitude of the anomalies are stronger and weaker, respectively.

The circulation changes associated with the thermodynamic response and with the impacts in T2m and precipitation over the Euro-Mediterranean region are discussed in the next section.

*c. Influence of the AMV-forced SST anomalies on the atmospheric circulation*

The Z500 response to the AMV is characterized by anticyclonic (cyclonic) anomalies centered over Greenland (northwestern Europe) and projects on NAO- pattern (Fig. 2ef). Previous studies focusing on the observed AMV impacts have suggested several mechanisms leading to a very similar Z500 response.

Based on observational analysis, Dong et al. (2012) suggest a local influence of the AMV on the NAO: during an AMV+ phase, the weakening of the SST meridional gradient over the subpolar gyre induces a southward shift of the westerly winds and the storm tracks whose fingerprint resembles the summer NAO-. This shift leads to similar temperature and precipitation anomalies to those shown in Fig. 2cd: enhanced precipitation over the UK and northwestern Europe and decreased precipitation over the Mediterranean basin (their Fig. 3). However, no significant anomalies in the summer storm tracks is detected in our idealised AMV experiments (not shown).

Several studies also suggest that the Z500 anomalies over Europe are part of a larger-scale pattern. Bladé et al. (2011) show that a negative (positive) phase of the NAO is associated with a Z500 tripolar structure, characterized by a Z500 dipole located over East Atlantic and northern Europe and an anomalous ridge (trough) over the Mediterranean basin. The authors interpret this tripole as part of a circumglobal wave-like pattern of anomalies over the Northern Hemisphere. A similar response is obtained in CNRM-CM5 and EC-Earth3P, which also indicates a wave-like pattern of Z500 anomalies in 2xAMV (Z500*, Fig. 8ab) with positive anomalies over the North Pacific, north-eastern America, Greenland the Mediterranean sea; and negative anomalies over western North America and northern Europe. This wave-like pattern is robust across the 3 amplitudes of the AMV (Fig. 9-10), with a stronger (weaker) amplitude in 3xAMV (1xAMV).

Several mechanisms implying the AMV have been proposed to explain this wave-like pattern. They are based on both the tropical and extra-tropical AMV-forced SST anomalies.
Ghosh et al. (2017, 2019) claim that the negative Z500 anomalies over western Europe result from a linear baroclinic response triggered by a diabatic heating over the northwestern Atlantic during an AMV+ phase (Gulev et al. 2013), inducing an easterly wave-like response. Lin et al. (2016) also suggest an influence of the AMV-forced extra-tropical SST on the multidecadal variability of the wave-like circumglobal pattern, but unlike Ghosh et al. (2017), they found in idealized experiments that the circulation anomalies over Europe are of a barotropic type.

On the other hand, a key role from tropical Atlantic in modulating the atmospheric circulation in the extra-tropics has also been mentioned by other studies. Wang et al. (2012) show that the observed multidecadal changes in the circumglobal teleconnection during the last century have been partially due to the West African monsoon. The latter is known to be enhanced during an AMV+ phase via the tropical Atlantic (Zhang and Delworth 2006; Martin et al. 2014), which is also the case in our experiments (Fig. 8cd). According to Wang et al. (2012), the African monsoon, acting as a heat source and coupled with the upper-level jet stream, generates a Rossby wave response, which may modulate the atmospheric circulation over the Euro-Atlantic region. Alternatively, Gaetani et al. (2011), Cassou et al. (2005), among others, suggest an other mechanism, which is based on a direct meridional overturning circulation between the tropics and the Mediterranean basin, triggered by the enhanced West African monsoon, and/or by the tropical Atlantic warming during an AMV+ phase.

Assessing (i) the relevance of each of the above-mentioned mechanisms in each model, and (ii) their sensitivity to the amplitude of the AMV requires further dedicated experiments and analyses, which are beyond the scope of this study. However, we provide here a qualitative estimation of the contribution of the tropical and extra-tropical part of the AMV on atmospheric circulation anomalies by using the additional twin ensemble experiments 1xAMV$_T$ and 1xAMV$_E$. Fig. 11 shows that the Z500* anomalies obtained over the Euro-Mediterranean region in the full pattern experiments for CNRM-CM5 are not due to an exclusive SST forcing from the tropics nor from the extra-tropics. The tripolar circulation anomalies over the Mediterranean basin in 1xAMV seem to be attributable to the tropical SST anomalies, with a similar northwestward tilted structure in 1xAMV$_T$, whereas the Z500* response obtained in 1xAMV$_E$ is weaker (Fig. 11abc). Conversely, the positive anomalies over north-eastern America seem to be attributable to an extra-tropical forcing, although the signals are hardly significant because of the weak amplitude of the SST anomalies.

The additivity hypothesis of the tropical and extra-tropical impacts on the atmospheric circulation is tested by comparing the sum of the 1xAMV$_E$ and 1xAMV$_T$ responses to the 1xAMV response. Fig. 11cd shows the comparison between the Z500* response in 1xAMV$_T$ + 1xAMV$_T$ compared to 1xAMV. The additivity assumption is not valid over the subpolar gyre and Central Europe where opposite-sign responses are obtained; however, the tripolar Z500* anomalies over the Mediterranean basin in 1xAMV$_T$ + 1xAMV$_T$, although slightly shifted, are similar to those obtained in the full pattern response. These results support the idea of the existence of non-linearities, *i.e.* a combined role of the tropical and the extra-tropical AMV-forced SST anomalies to explain the atmospheric response over the Euro-Atlantic region, even if, again, the amplitude of the SST forcing is too weak to firmly conclude.

The spatial patterns of the T2m, HW and Z500 responses shown in Fig. 2-4 indicate a progressive reinforcement of the anomalies with the amplitude of the AMV in both models. The next section aims at quantifying the degree of linearity between the atmospheric response and the amplitude of the AMV-forced SST anomalies.

*d. Linearity between the AMV-forced SST anomalies and the HW response over the Mediterranean basin*

The intensity of the T2m response seems to be a linear function of the amplitude of the AMV for both models (Fig. 12a). A linear relationship is also obtained for the HW response, but unlike the T2m responses, the slope is, according to a t-test, significantly greater for EC-Earth3P than for CNRM-CM5 (Fig. 12ab). The intercept is also higher for EC-Earth3P, and denotes that the climatological number of HW days is higher in this model. This discrepancy between the two models in the AMV/HW relationships can be due to two reasons: (i) different intrinsic model differences leading to different responses, and (ii) different external forcing background states (1950 for EC-Earth3P and 1985 for CNRM-CM5). The first one could be associated with different climatologies between the two models impacting the HW response: *e.g.* a weaker climatological soil moisture in EC-Earth3P than in CNRM-CM5 during summer. Assessing the contribution of this feature would require further experiments, which are beyond the scope of the study.

The sensitivity of the response to the external forcing background is evaluated with CNRM-CM5, for which additional simulations were performed with an 1850 external forcing background (Qasmi et al. 2019). These simulations are similar to those used in this paper except that they are initialized from the conditions of the CMIP5 pre-industrial control simulation of CNRM-CM5, in which external forcings are maintained at the 1850 level. The comparison between these two types of experiments, named CNRM-CM5(1850) and CNRM-CM5(1985), shows that the mean background state has no clear influence on the AMV/T2m linear relationship (Fig. 12c): although the T2m response in the CNRM-CM5(1850) simulation is weaker with a lower intercept of ~0.1°C, the same slopes are obtained in for both mean background states. A similar result is obtained for the HW response, for which CNRM-CM5(1985) and CNRM-CM5(1850) are almost indistinguishable from each other. Note that the linearity of the T2m/HW response does not necessarily imply a linearity of the mechanisms leading to this response. The precise origins of the latter, discussed in section 3.3, as well as their sensitivity to the amplitude of the AMV still remain to be clearly identified. Note also the larger spread of the T2m and HW anomalies in CNRM-CM5(1850) than in CNRM-CM5(1985). The reduction of the spread with the 1985 background may correspond to a stronger signal-to-noise ratio in the latter, which could be interpreted as the impact of the anthropogenic forcing, which tend to reduce the atmospheric noise over the Mediterranean basin (Bengtsson et al. 2006). Assuming that the sensitivity to the external forcing background is the same in EC-Earth3P, *i.e.* that the interaction between AMV and the external forcing background is not model-dependent, the differences between the two models in the AMV/HW linear relationships may depend on their respective intrinsic biases.

## 4. Conclusions and discussion

The impacts of the AMV on the European summer climate are studied in two climate models, EC-Earth3P and CNRM-CM5, using idealized experiments in which the North Atlantic SST is restored towards anomalies characteristic of the observed AMV. To estimate the sensitivity of the atmospheric response to the amplitude of the AMV, three ensembles of simulations are performed , corresponding to one, two and three standard deviations of the observed AMV. Both climate models are coherent in the surface temperature and precipitation responses and show that during a positive phase of AMV an increase (decrease) in T2m and a decrease (increase) in precipitation are obtained over the Mediterranean basin (northern half of Europe). These results are consistent with previous studies based on a similar experimental protocol (Ruprich-Robert et al. 2017), and observations (Sutton and Dong 2012, their Fig. 2 and 3). For an observed AMV amplitude of two standard deviations, extreme temperature events increase by up to 50% in the number of days of HW over several regions of the Mediterranean basin, especially over its eastern part, where both models exhibit the most robust response. The associated mid-troposphere response is characterized by an anomalous trough (ridge) over Northern Europe (the Mediterranean Sea), which is coherent with the responses in T2m, HW and precipitation.

A decrease in total cloud cover over the Mediterranean basin during an AMV+ phase induces an increase in downward shortwave radiation, which contributes to the surface warming and to the increase in the number of HW days. In addition, as a response to the decrease in precipitation, evapotranspiration is reduced, contributing to a surface warming. Finally, the advection by the climatological flow of warmer air mass from North Atlantic and the Mediterranean sea provides an additional warming over northwestern Africa, the Iberian Peninsula and the Levant.

Different mechanisms, which have been addressed in previous studies, could explain the changes in the atmospheric circulation leading to the cloud cover anomalies and the associated responses in T2m, HW and precipitation over the Euro-Mediterranean region. The Z500 anomalies over the Euro-Atlantic region are found to be part of a circumglobal stationary wave in the Northern Hemisphere which is more excited during a positive phase of the AMV. Here we argue that the atmospheric circulation anomalies over the North Atlantic and *a fortiori* over the northern hemisphere, may be the result from numerous non-exclusive sources, that in turn can be classified into extra-tropical and tropical sources.

An extra-tropical warming over the North Atlantic SST during an AMV+ phase could enhance diabatic heating over the midlatitudes inducing an anomalous eastward-propagating wave. The midlatitude circulation could also be modulated by a warmer than normal Tropical Atlantic that may (i) enhance the diabatic heating causing an anomalous wave activity in the upper troposphere over the Euro-Atlantic region and/or (ii) trigger a direct meridional atmospheric cell between the tropics and the Mediterranean basin. Although not addressed in this study, remote effects of AMV on the adjacent oceanic basins (Ruprich-Robert et al. 2017) may also impact the geopotential response via (i) the tropical Pacific through divergence anomalies (O'Reilly et al. 2018), and (ii) the Indian ocean and the Asian monsoon through the eastward propagation of Rossby waves (Rodwell and Hoskins, 2001; Cherchi et al. 2014), but a deeper analysis goes beyond the scope of the present paper. Identifying the relevant mechanisms, assessing their respective contribution in each model, and quantifying their sensitivity to the amplitude of the AMV constitute appealing perspectives within the DCPP-C initiative.

A linear relationship between the amplitude of the AMV and the HW/T2m response is found in both models, with a stronger AMV/HW relation in EC-Earth3P due to intrinsic model differences between the two models. One caveat must be raised regarding the interpretation of this linear relationship. Since the restoring coefficient is fixed, the SST restoring tends to be more efficient in the tropics than in the extra-tropics due to the different mixed layer depths (MLD) between these two regions. SSTs are better constrained in the tropics where the ocean is more stratified, than in the extra-tropics where the mixed layer is thicker. Therefore, more weight may be given to the tropics than the extra-tropics, potentially leading to (i) an underestimation of the mechanisms of teleconnection from this region, (ii) a different

HW response over Europe, and (iii) a breaking of the linear relation between SST and HW. As done by Ortega et al. (2017), additional experiments in which the restoring coefficient evolves as a function of the MLD have been computed with CNRM-CM5 and EC-Earth3P. Preliminary results indicate that the responses are unchanged.

The similarity between models and observations suggests that the observed T2m anomalies over Europe result from a forcing by the AMV rather than being a consequence of the sole atmospheric internal variability. Estimating the HW sensitivity to the AMV over the Mediterranean basin in observations would contribute to assess the robustness of our findings.

A last interesting perspective would also be to evaluate the sensitivity of the response to warmer mean background states. We found that this feature has little influence in CNRM-CM5, but there is no guarantee that for a considerably warmer climate, typically characteristic of the mid- to late-21st century, the level of teleconnectivity remains the same given the future abrupt climate changes that could occur depending on anthropogenic emission scenarios. This would provide an estimate of the risks associated with HW few decades ahead.

**Acknowledgements:**

This work was supported by a grant from Electricité de France (EDF), by the French National Research Agency (ANR) in the framework of the MORDICUS project (Grant Agreement ANR-13-SENV-731 0002), by the European Union's Horizon 2020 Research and Innovation Programme in the framework (i) of the PRIMAVERA project (Grant Agreement 641727) and (ii) of the Marie Skłodowska-Curie grant INADEC (Grant agreement 800154). The figures were produced with the NCAR Command Language Software (http://dx.doi.org/10.5065/D6WD3XH5).

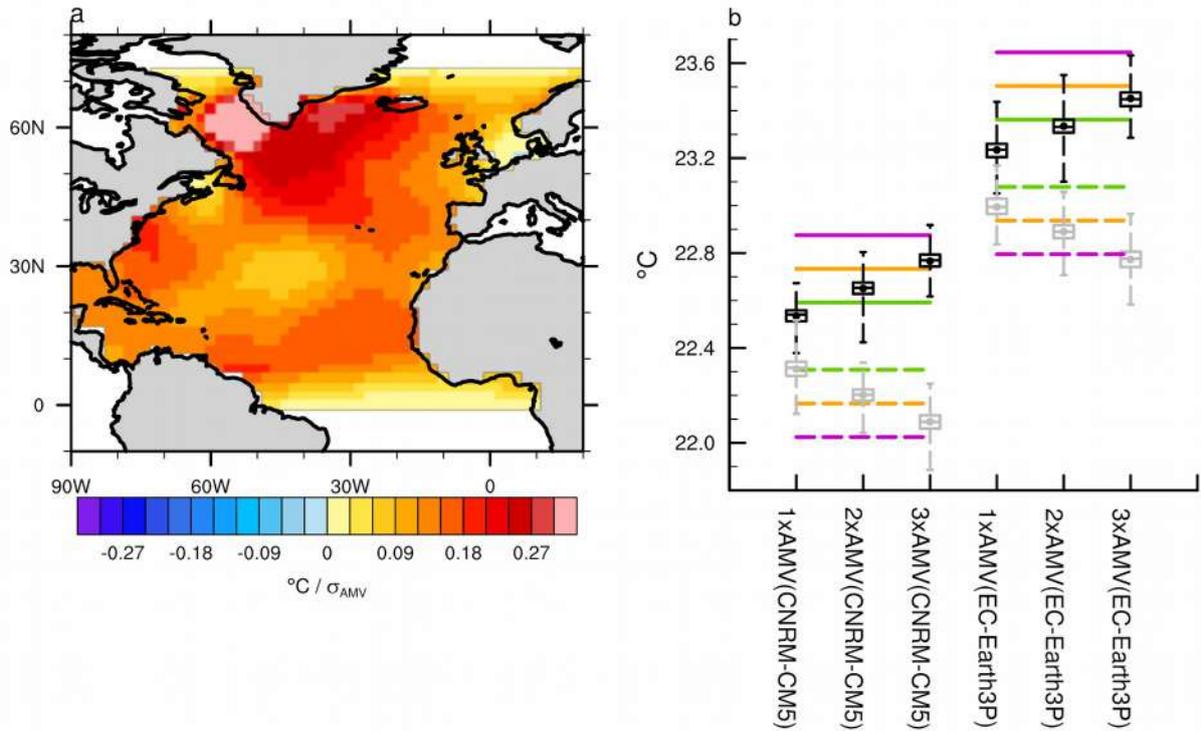

Fig. 1: (a) Anomalous SST pattern used for restoring and taken from input4MIPs archive (units are °C/σ(AMV), shading interval is every 0.03 °C). (b) Simulated raw annual SST averaged over the North Atlantic restored sector for AMV+ (black) and AMV- (gray) experiments for CNRM-CM5 and EC-Earth3P. Each boxplot stands for the distribution obtained from 250 years for each ensemble (25-members × 10-years). The top (bottom) of the box represents the first (last) tercile of the distribution and the upper (lower) whisker represents the first (ninth) decile. Dots and inside-line stand for the mean and the median of the distribution, respectively. The green, orange and magenta horizontal lines show the SST targets for each model for the 1xAMV, 2xAMV and 3xAMV ensembles corresponding to 1, 2 and 3 standard deviations of the observed AMV index, respectively. Solid and dashed stands respectively for AMV+ and AMV- experiments.

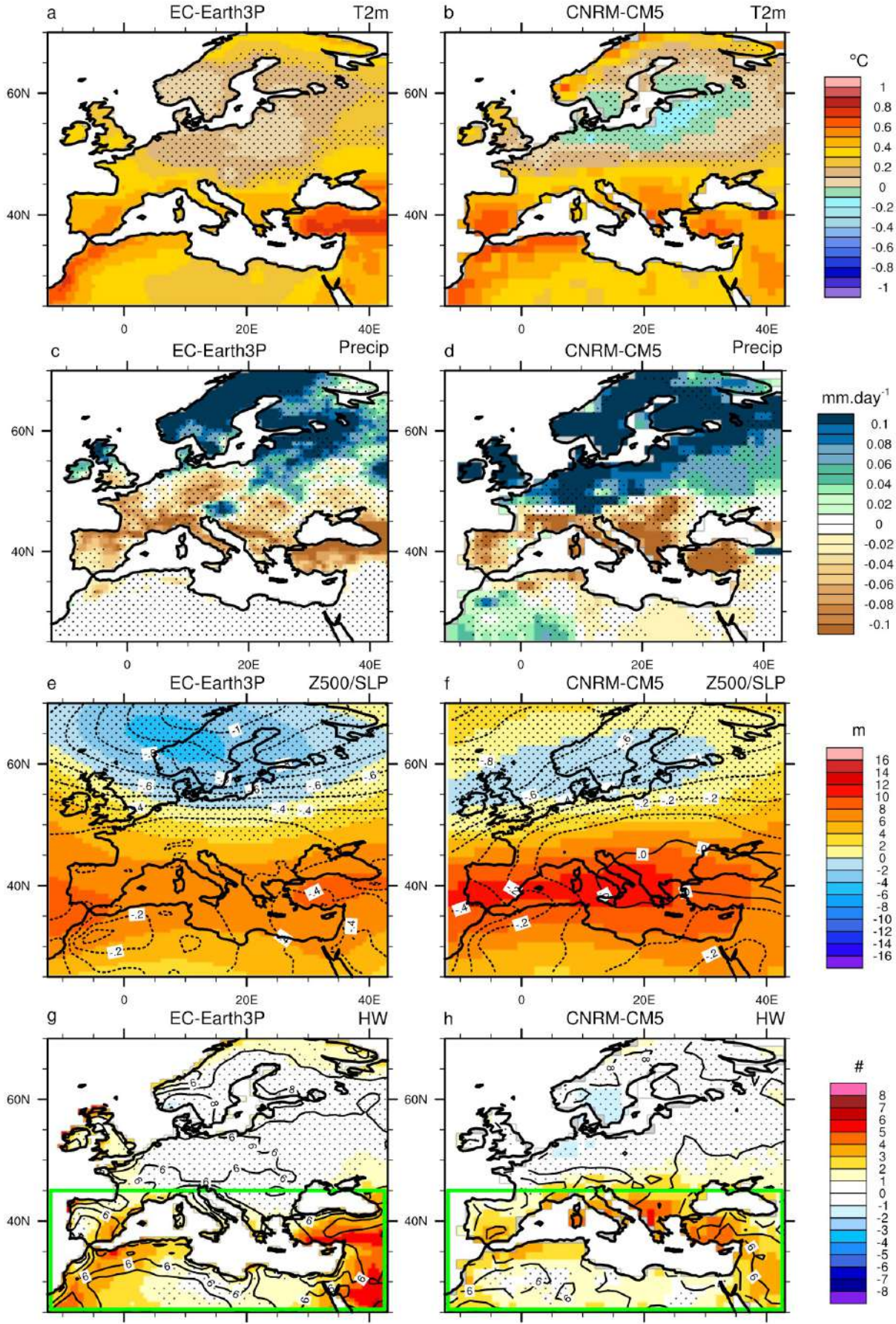

Fig. 2: AMV-forced anomalies for June-August seasonal mean for T2m (ab, shading interval is 0.1°C), precipitation (cd, shading interval is 0.01 mm.day-1), SLP (ef, contour interval is 0.1 hPa and the thicker black contour stands for the zero line) superimposed on Z500 (shading interval is 2 m), and number of HW days (gh, shading interval is 1 day) for EC-Earth3P (left) and CNRM-CM5 (right). Stippling indicates regions that are below the 95% confidence level of statistical significance based on two-sided Student's t-test.

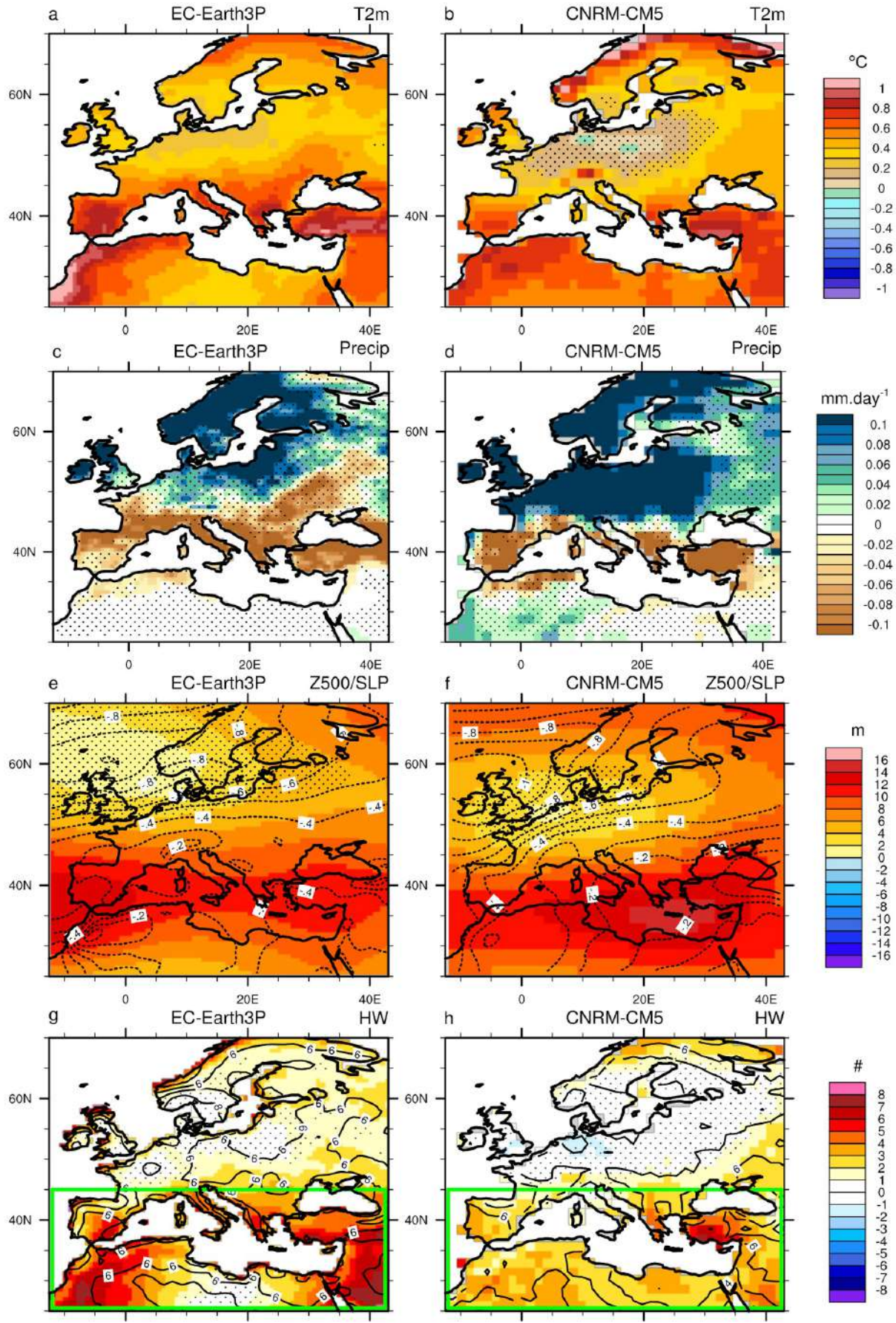

Fig. 3: Same as Fig. 2 but for 3xAMV.

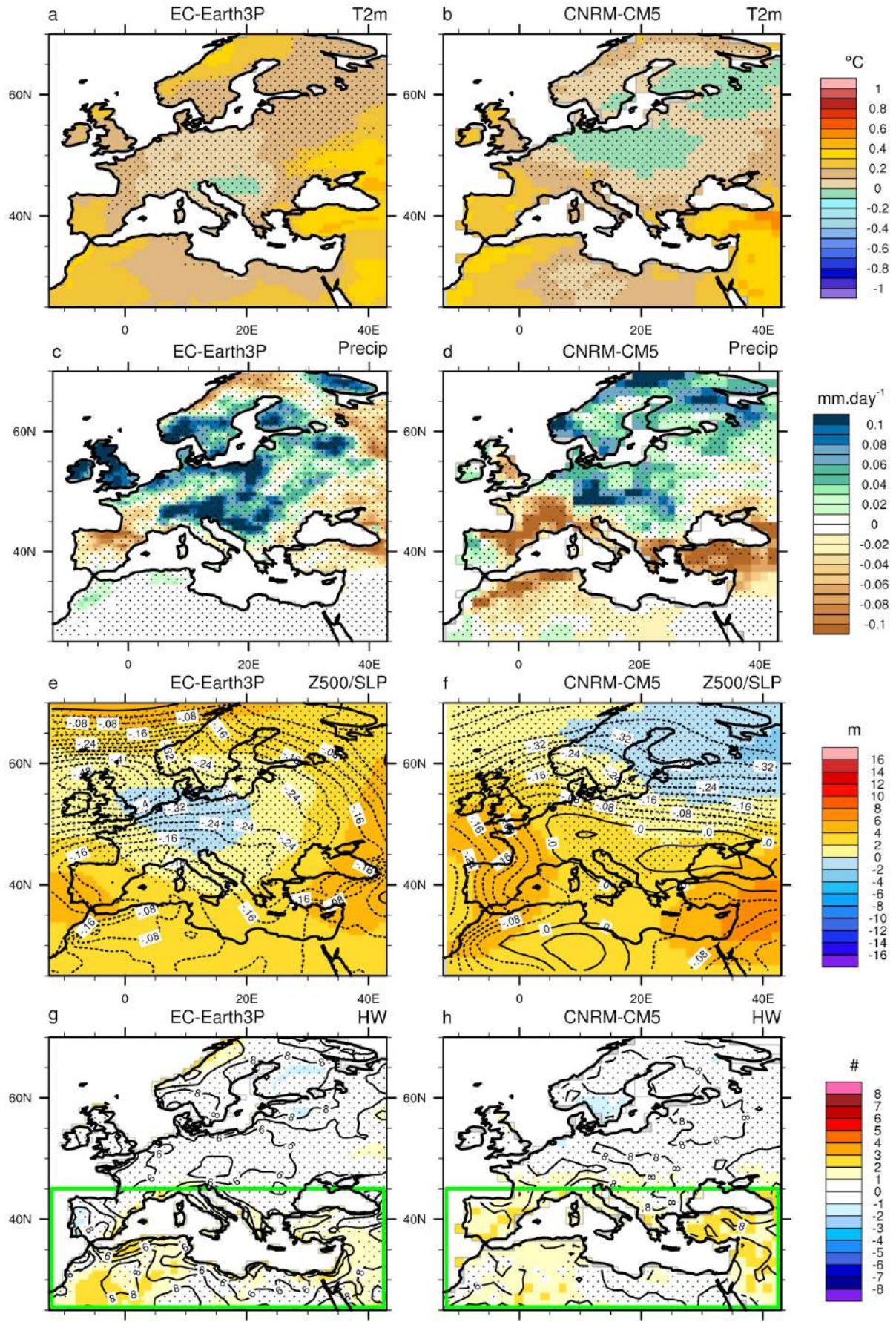

Fig. 4: Same as Fig. 2 but for 1xAMV.

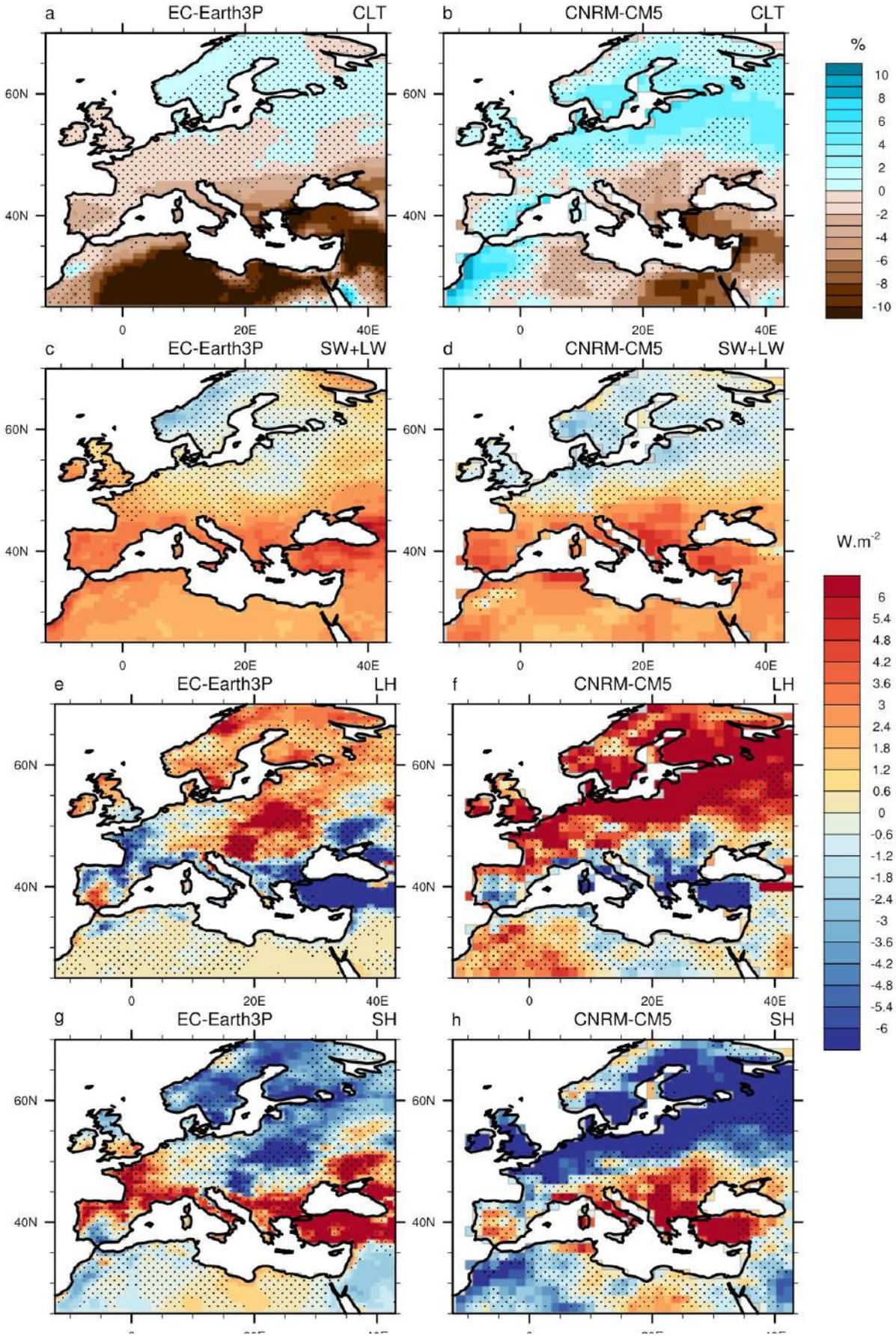

Fig. 5: AMV-forced anomalies for June-August seasonal mean for total cloud cover (ab, shading interval is 1%), downward LW and SW radiation at surface (cd, shading interval is 0.6 W m$^{-2}$), latent heat at surface (ef), and sensible heat (gh) for EC-Earth3P (left) and CNRM-CM5 (right). Positive values represent heat transfer towards the surface. Stippling indicates regions that are below the 95% confidence level of statistical significance based on two-sided Student's t-test.

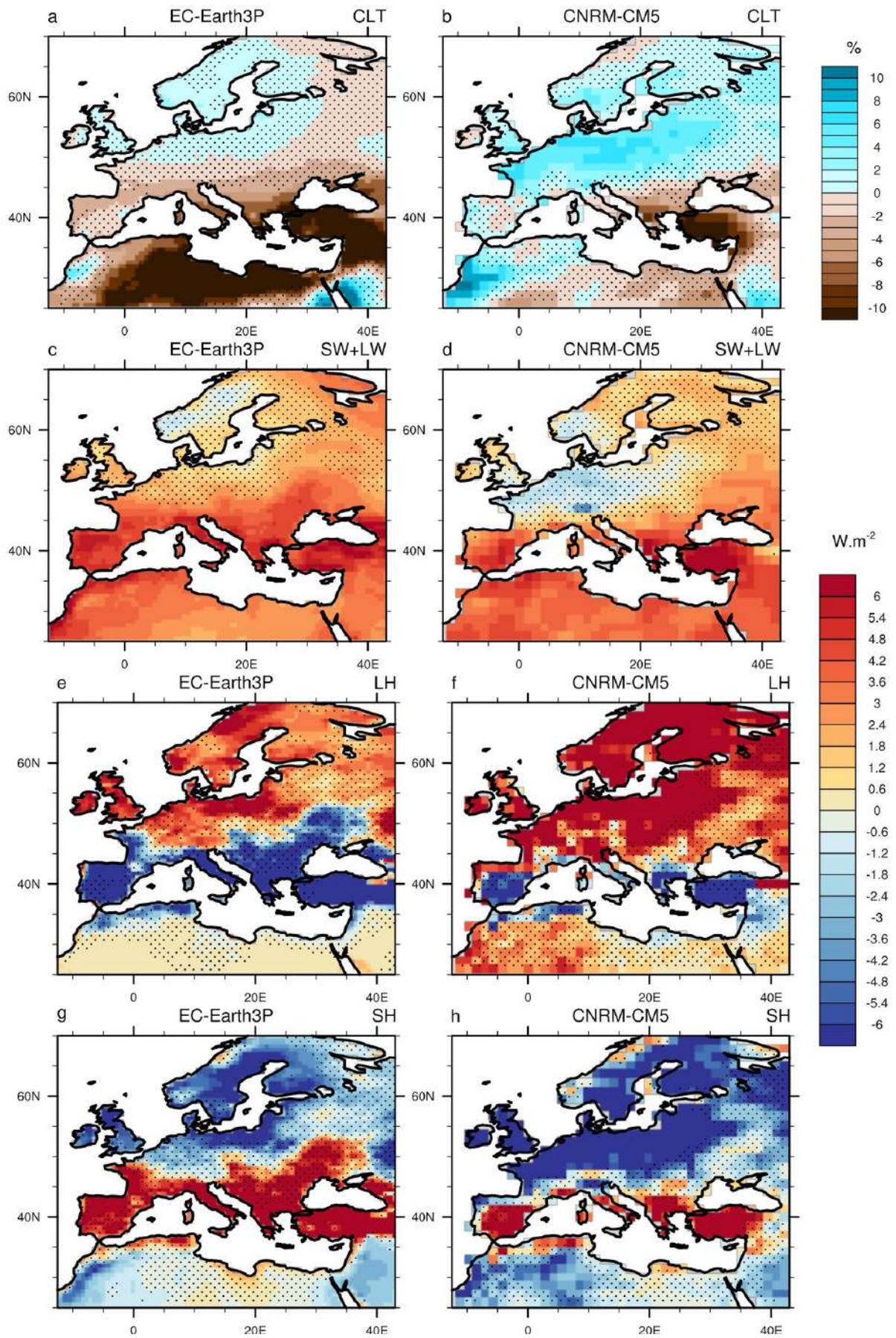

Fig. 6: Same as Fig. 5 but for 3xAMV.

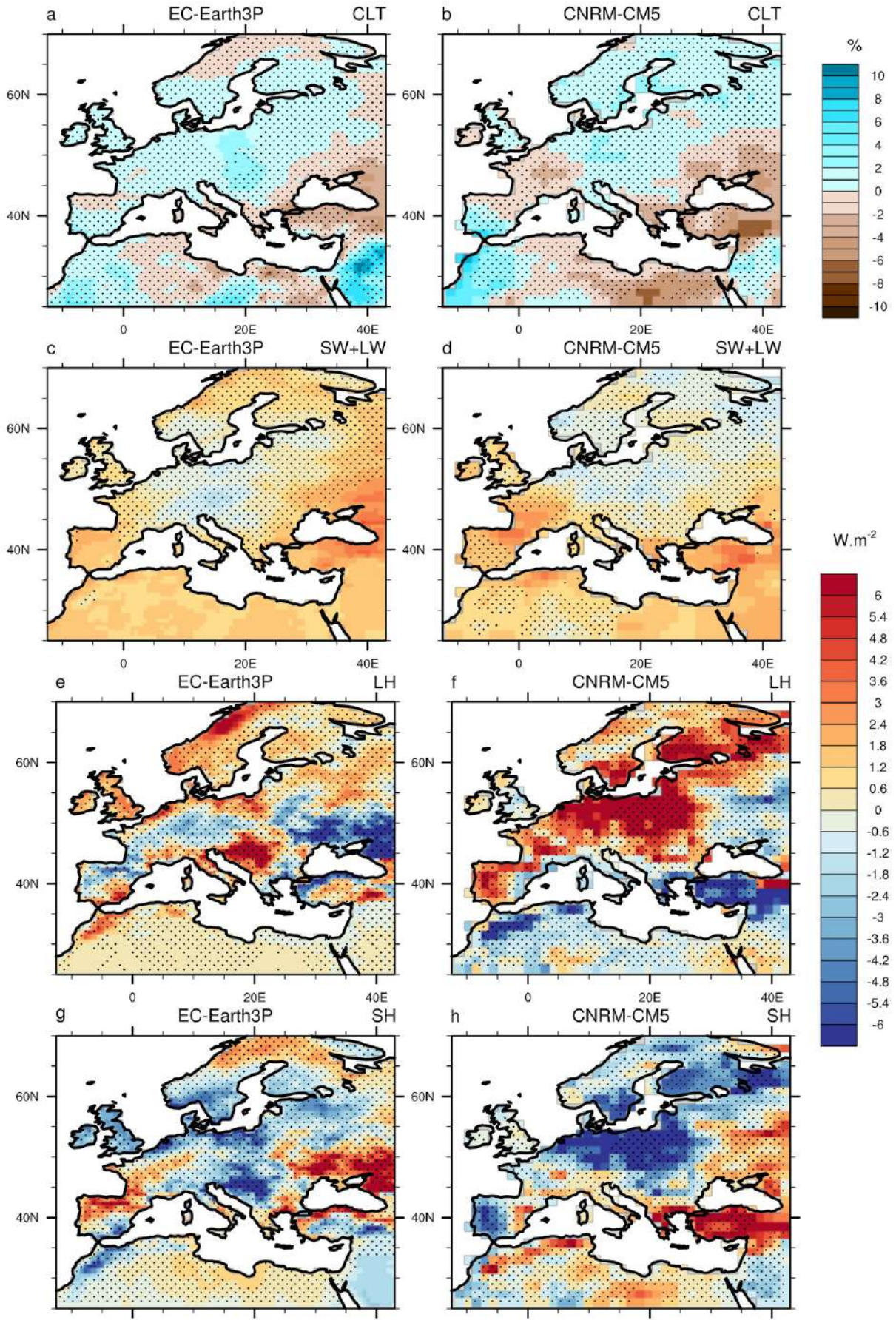

Fig. 7: Same as Fig. 5 but for 1xAMV.

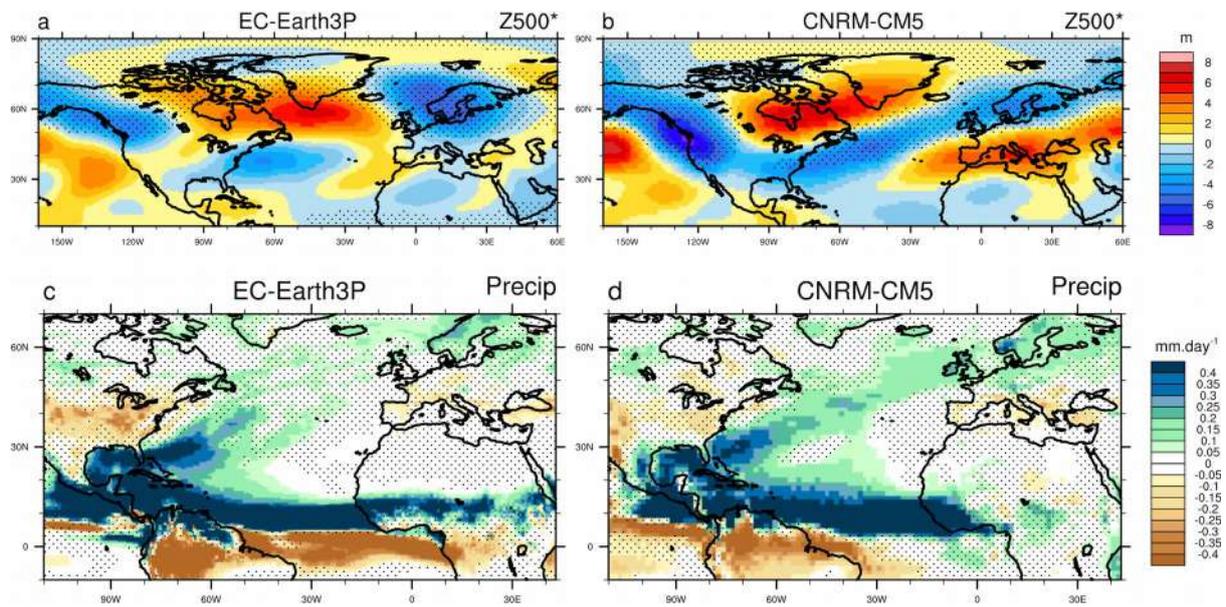

Fig. 8: AMV-forced anomalies for June-August seasonal mean for Z500* (ab, shading interval is 2 m), precipitation (cd, shading interval is 0.01 mm.day-1) for EC-Earth3P (left) and CNRM-CM5 (right). Note that Z500 zonal means have been retrieved to account for the mean dilatation of the atmosphere due to the artificial heat source introduced in the model in the idealized experiments via the flux restoring term. Stippling indicates regions that are below the 95% confidence level of statistical significance based on two-sided Student's t-test.

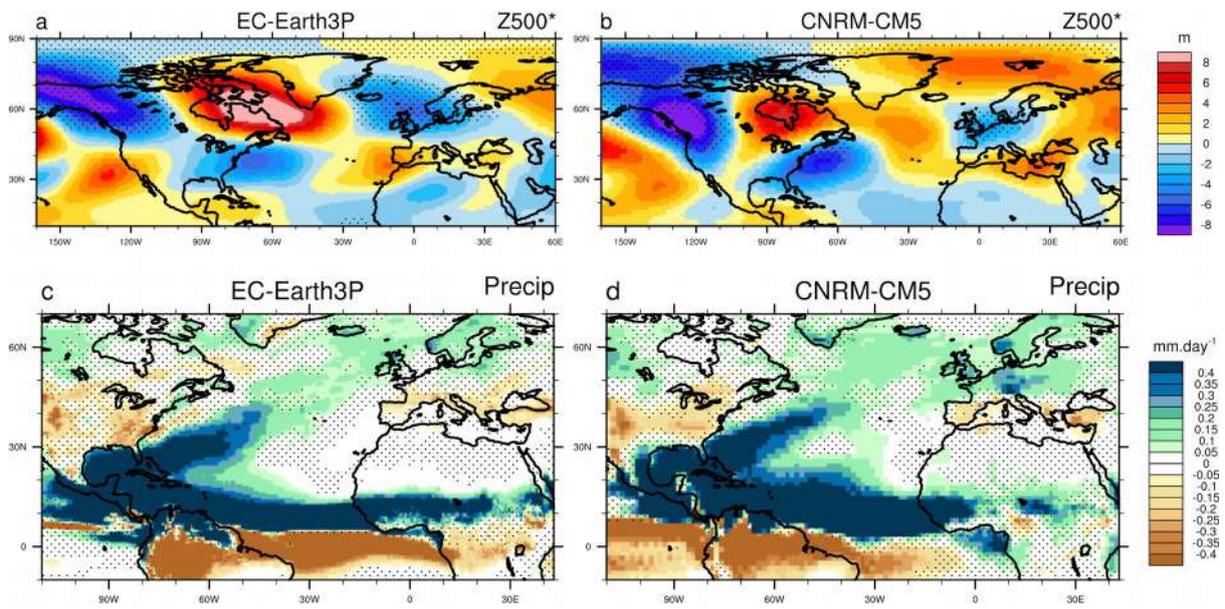

Fig. 9: Same as Fig. 8 but for 3xAMV.

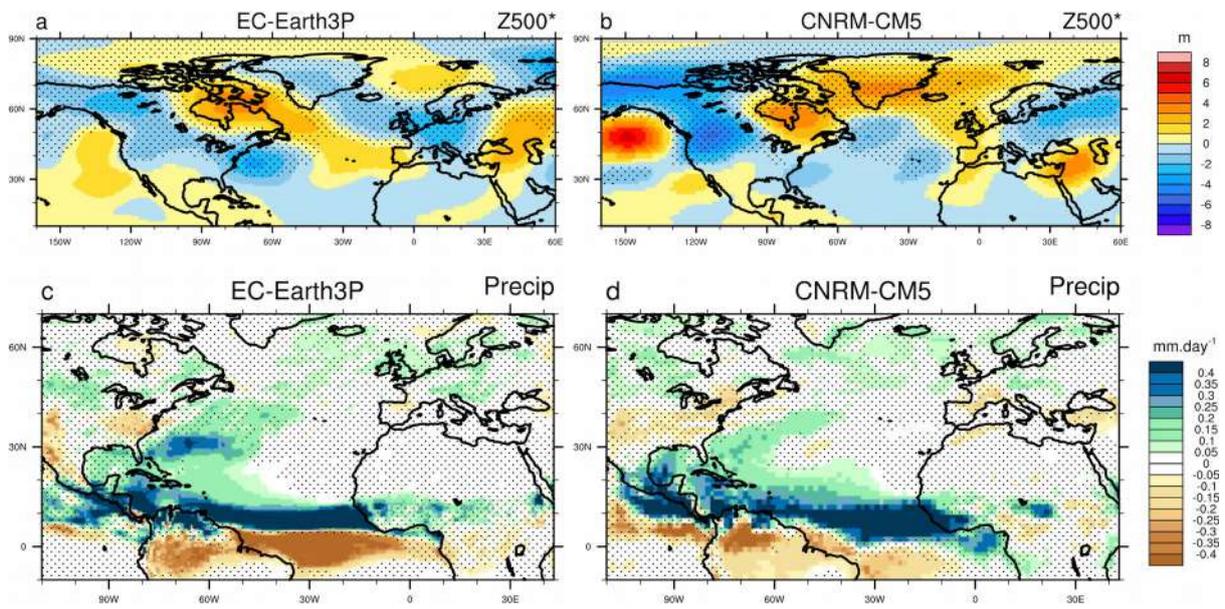

Fig. 10: Same as Fig. 8 but for 1xAMV.

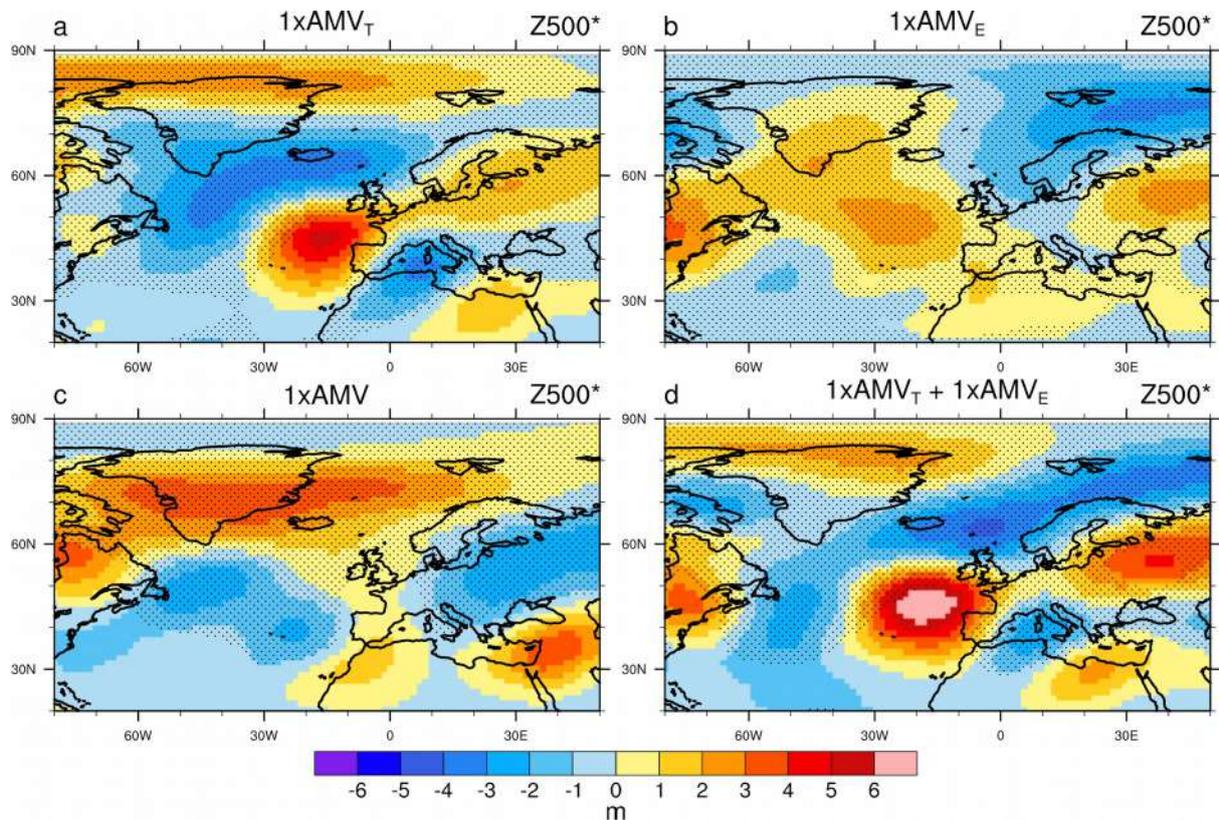

Fig. 11: AMV-forced anomalies for June-August seasonal mean for Z500* (shading interval is 1 m), for CNRM-CM5 for $1xAMV_T$ (a), $1xAMV_E$ (b), $1xAMV$ (c) and the sum $1xAMV_T + 1xAMV_T$ (d). Stippling indicates regions that are below the 95% confidence level of statistical significance based on two-sided Student's t-test.

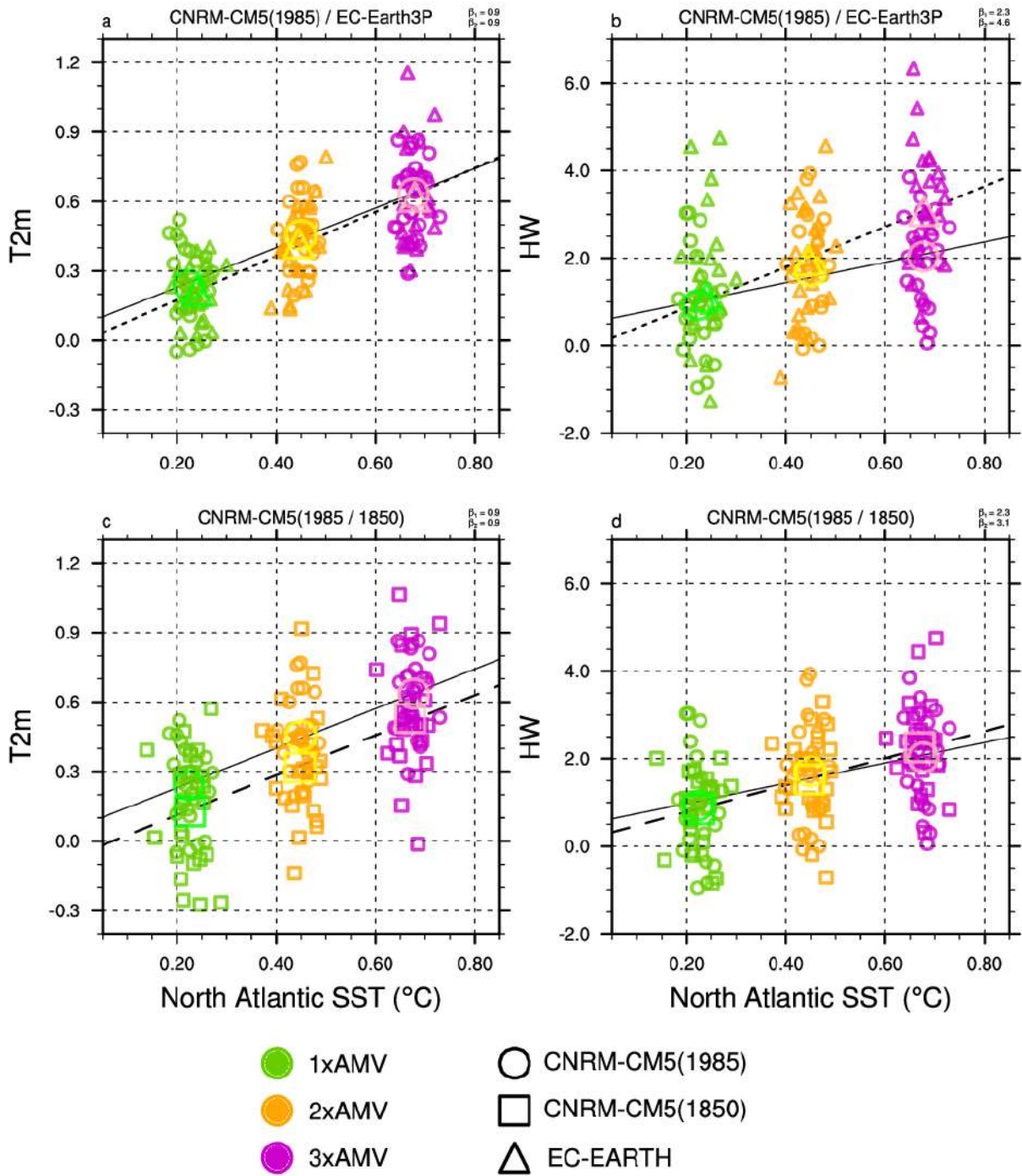

Fig. 12: Spatial average of AMV-forced anomalies for June-August seasonal mean of North Atlantic SST (0°-60°N) versus T2m (left column) and number of HW days (right column) over the Mediterranean basin (green domain over Fig. 2gh) for CNRM-CM5(1985) and EC-Earth3P (top line), and for CNRM-CM5(1985) and CNRM-CM5(1850) (bottom line) for 1xAMV (green), 2xAMV (orange) and 3xAMV (magenta). The small dots and squares represent the 10-yr mean response of each member and the big dot/square stands for the ensemble mean. The slope β obtained from the linear regression between the T2m/HW and the SST anomalies distributions from all the experiments is given in the upper right title of each panel.